# Mobile Services and ICT4D
## *To the Network Economy - Bridging the Digital Divide, Ethiopia's Case*

Naod Duga[#*1], Henok Getachew[#2]

[*] Department of Electrical and Computer Engineering, Addis Ababa University
[1] naod@ece.aau.edu.et

[#] Emopia Information and Communications Engineering
[2] henok.getachew@emopia.com

Addis Ababa, Ethiopia

*Abstract*— This paper presents a development paradigm for Ethiopia, based on appropriate services and innovative use of mobile communications technologies via applications tailored for sectors like business, finance, healthcare, governance, education and infotainment. The experience of other developing countries like India and Kenya is cited so as to adapt those to the Ethiopian context. Notable application areas in the aforementioned sectors have been outlined. The ETC 'next generation network' is taken into consideration, with an emphasis on mobile service offering by the Telco itself and/or third party service providers. In addition, enabling technologies like mobile internet, location-based systems, open interfaces to large telecom networks, specifically service-oriented architecture (SOA), Parlay/JAIN and the like are discussed. The paper points out possible endeavors by such stakeholders like: telecom agencies and network operators; businesses, government and NGOs; entrepreneurs and innovators; technology companies and professionals; as well as researchers and academic institutions. ICT4D through mobile services and their role in bridging the digital divide by building a virtual 'network economy' is presented.

## I. INTRODUCTION

The significance of mobile telecommunications services beyond basic person-to-person voice communication can't be overemphasized. The now flourishing telecom networks and feature-rich mobile handsets coupled with innovative services and application are perhaps the next 'killer' solution providers for what many people (in developing nations, especially) could look up on. We try to strengthen our argument with a pragmatic discussion of what researchers and key figures refer to as, "mobiles bridging the digital divide"[6], "the mobile economy"[2], "the convergence of computing and communications"[4] in "the information age"[1] while taking into account a developing nations' perspective.

Let's define some terms first. According to Philip Kotler (1998), a 'service' can be defined as "any act or performance that one party can offer to another that is essentially intangible and does not result in the ownership of anything" (p. 477). In mobile telecommunications, the service has traditionally been a subscription to the voice services, that is, making phone calls over mobile networks. Now, though, mobile services have gone far beyond basic mobile telephony to services relying on data over telecom networks as in messaging & mobile internet and a diverse range of vertical applications on top of enabling technologies brought by the telecom/IT convergence.

In the past, a nation's wealth was largely a function of its endowment of natural resources, its labor force and its accumulated capital base. The 'new' wealth of nations places increasing emphasis on knowledge in its various embodied forms. ICTs permit information and knowledge to expand in quantity and accessibility. In many fields there is the possibility of unprecedented new tools for development. In such fields as agriculture, health, education, human resources and environmental management, the consequences could be revolutionary. ICTs have enormous potential especially for developing countries [2] and that's what ICT4D (ICT for Development) stands for.

The role played by mobile networks, services and applications in developing countries' endeavours is certainly huge. We shall take Ethiopia as an example but the discussions would surely apply for others as well. It's our belief that mobiles are the next big thing, of course, if used appropriately in a timely and innovative basis. They inevitably are key players in ICT4D endeavours in developing nations.

In Section II of this paper, we discuss the challenges and opportunities in mobile services and ICT4D by introducing the so-called 'mobile economy' and then shade light on technology and infrastructure trends in the mobile telecom services. We further cite exemplary cases of mobile services from other countries. We finish the section with a discussion of noteworthy mobile service use cases in different scenarios.

Section III focuses on Ethiopia, building up on the status quo of mobile networks, services and applications together with reference to stakeholders in the mobile arena. With the network in mind, we cite enabling technologies and platforms for mobile network services and applications. The need to open up networks for third parties with Open APIs such as Parlay/JAIN will also be discussed. We finish the section with notes on mobile services, from concept to design then to deployment and quality assurance. We assert the complexity of the mobile service domain [2] and point out directions towards creation of winning mobile services.

We wrap up our paper with a discussion of the role of the different stakeholders in Section IV. A conclusion section is also included.

## II. CHALLENGES AND OPPORTUNITIES

Think 6.7 billion people, bring in 4 billion (and counting) mobile phones immersed in thousands of petabytes of data (or information, if we may) on the Internet, introduce innovative services like location based systems, market information, mobile banking and social networking in addition to everyday voice and data communications to end users as diverse as a Masai hunter or an Ethiopian farmer to a Fortune 500 executive. What would we get?

### A. The Mobile Economy

The mobile phone has emerged as an indispensable communications and computing machine in our fingertips. It has enabled the convergence of information technology and telecommunications systems, being a platform for a range of services from basic telephony to data services for people from around the world. Advances in consumer electronics and software technologies is bringing handsets that are increasingly usable, cheaper and full of features to consumers, allowing them to communicate, learn, socialize and do business in an increasingly connected world.

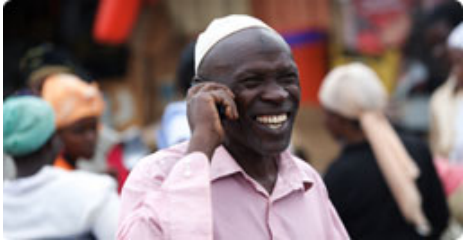

Figure 1 - Africa is among the fastest growing mobile telecom markets.

The number of mobile phone subscribers is expected to hit 4.5 billion by 2012 globally and this, for the most part, is attributed to emerging markets in Asia, Latin America and Africa. There are more than 300 million mobile subscribers in Africa alone, according to a report by the ITU (International Telecommunications Union).

The global telecom market is now worth $4 trillion. By 2011, global telecommunications revenue is estimated to hit close to $5 trillion, with the growing demand for high-volume data applications driving both business and consumer markets, reports TIA, the Telecommunications Industry Association.

The mobile telecom industry, in its relation to other economic sectors like business, agriculture, health and education, is creating a virtual 'economy' of its own – being at the heart of communications, information provision, entertainment and commerce, among others.

### B. Telecom Technology and Infrastructure Trends

Changes in telecommunications are impacting all types of user group, which include business users, travelling users, small and home offices, and residential users.

Internet access, special data services on top of traditional voice networks, voice services on top of data networks as well as multimedia services are among those made possible through the fast growing global telecommunications market.

Telecom networks are exhibiting increasing complexity. Telecom service providers deploy emerging services based on a mixture of infrastructures such as mobile and wireless, xDSL and fiber as an overlay. [15]

One other trend is deregulation and privatization of the telecom network. This has led to a number of operators as well as third parties like ISVs and content providers serving the customer base.

Communication and computing convergence is one other trend worth pointing out.

The old adage 'The customer is king' is in the same way applicable for telecom customers and it will possibly be so in the foreseeable future as a result of competition between operators. The subscriber expects network service performance reliability, high QoS, flexible package subscription and, of course, cheaper bills.

Of special interest to our discussion, the following points are worth mentioning regarding mobile networks, devices, content, services and applications.

#### 1) Mobile Network Technologies:

In most countries, current second generation mobile networks based on GSM and CDMA are evolving to 3G and beyond with the introduction of WCDMA and CDMA2000. 2.5G services like GPRS and EDGE do compensate for the gap in many a place, as the third generation networks are seeing varied penetration and adoption rates.

#### 2) Mobile Devices:

Of all the components in a wireless network system, the mobile device gets the most attention. End users can only interact with the system through mobile devices. Mobile devices feature a low-power transceiver that is typically designed to transmit voice and data. The mobile device connects to the network infrastructure wirelessly through an available access point or base station. [15]

Notable operating systems include Symbian OS, Google Android and Windows Mobile, the Apple iPhone OS, Mobile Linux, RIM and Palm OS. Java, with the largest install base, has emerged as the software enabler for media, messaging, mapping as well as location-based services.

Processing power is ever increasing – for the CPU and the DSPs handling the low level communications. In addition, separate enabling chips for ringtone and sound, camera sensor, mobile graphics are being bundled in high-end mobile devices.

Though processing power continues to follow Moore's law, doubling every 18 months, batteries have continued to be a limiting factor and this, though evolving, is not likely to change.

Add-on features like image and video capture, connectivity (3G, Bluetooth, Wi-Fi, etc.), display and multimedia experience continue to be a key selling point for mobile devices as many mobile services make good use of these features as in GUIs, imaging, browsing and gaming.

#### 3) Content and Service Delivery:

Web content, organizations' and business' proprietary databases, market information, news updates and media like ringtones, music, audio/video educational content, etc., are among the content upon which a spectrum of mobile services can be built.

SMS is the predominant service bearer for such applications where bandwidth is limited or where textual content suffices. MMS is also catching up, especially with entertainment and media related services. Text as well as binary content of limited size (160 bytes, to be exact) can be sent and received via SMS. MMS allows for rich text, image, audio and video content with operator-specific message size, usually between 30 and 100Kb.

Voice also can be used as a service delivery platform as in IVR (Interactive Voice Response), for example, where menu-based services can be accessed via DTMF tones generated by a user pressing handset keypad.

The mobile web (equally referring to WAP or the now common internet via mobile IP) is a significant step forward for service provision on mobile devices.

*C. Experience of Other Countries*

Rather luxe-looking mobile services (from an economic and social development perspective, at least) like mobile entertainment, location-aware services and social networking are now found in abundance in the developed world. Established and emerging service providers as well as different organizations do utilize widely available bearer networks available (in Japan, Korea, the EU and USA, for instance) to provide a spectrum of mobile value-added services to mostly literate and financially able customers.

It will be appropriate to cite noteworthy efforts from developing nations like India and African counterparts like Kenya and Uganda – as their experience in using mobile services towards development can easily be adapted to the Ethiopian context and can shade light on the possible impact innovative mobile services for different scenarios might have in solving the country's development problems.

In Kerala, a state located in south-western India, farmers adopt SMS service to know rubber prices. Tata Teleservices, a major Telco in India, empowers Indian farmers use mobile phones to control irrigation. 'The cell phone has been crucial in binding India together', reported the *New York Times*, 'India now adds more cellphone connections than anyplace else, with 15.6 million in March [2009] alone. The cost of calling is among the lowest in the world. And the device plays a larger-than-life role there — more so, it seems, than in the wealthy countries where it was invented.' Mobile communication is revolutionizing economic and social development in rural India in sectors like microcommerce, finance, healthcare, governance, education and infotainment.

Not so far from India, disaster-prone Bangladesh trialled cell phone alerts – tens of thousands of mobile users in the country's flood and cyclone-prone areas will receive advance warning of an impending natural disaster through a flashing alert on their cell phones. Another popular service, CellBazaar, offers a virtual marketplace for the 20 million mobile-phone subscribers of Bangladesh's GrameenPhone Ltd. , enabling customers to trade a variety of items.

Throughout Africa, successful cases of mobile services have been in the news. In Tanzania, for example, mobile phones have hugely benefited the fishing community. Cell phones have had positive impacts in grain markets in Niger. Despite the absence of electricity, alternative power solutions have been harnessed, including personal movement and battery packs to charge cell phones. Text messages are playing their part to tackle HIV in South Africa and to fight malnutrition and save lives in Malawi. In Kenya and other developing countries, SMS scheme is being used to tap work force – allowing software localization to be done by natives who earn money in return, via mobile banking services like Kenya's M-PESA.

In addition to several vertical application cases of mobile services [6], large scale initiatives and service offering like mobile payment would inevitably have bigger impacts. Such would be possible in cases where governments, telecom operators, network vendors, NGOs and ISVs cooperate for a common vision – bringing about socioeconomic development.

*D. Notable Mobile Service Use Cases*

Entertainment, customer care, business applications, finance, security, shopping, advertising, information provision, etc are among the use cases where mobile networks, services and applications are employed.

1) *Mobile Banking and Mobile Payment:* Mobile banking involves performing balance checks, account transactions, payments etc. via a mobile device such as a mobile phone. Also known as m-banking, it is most often performed via SMS or mobile Internet but can also use special programs installed to the mobile device. Challenges include handset interoperability, security, scalability & reliability, application distribution and personalization.

In mobile payment, a consumer uses a mobile phone to pay for different services as well as soft or hard goods instead of paying with cash, check or credit cards. Primary models include direct billing (operator or mobile), contactless NFC (Near Field Communication) and online payment using such popular services like PayPal, Amazon Payments and Google Checkout.

2) *Mobile Commerce:* Mobile Commerce is defined as 'any transaction, involving the transfer of ownership or rights to use goods and services, which is initiated and/or completed by using mobile access to computer-mediated networks with the help of an electronic device.' [10]

3) *Mobiles in Health:* m-Health can be defined as the use of mobile communications and network technologies for healthcare [11]. This might include education and awareness, data collection (for public health or clinical domains), remote monitoring, communication and training for healthcare workers, disease surveillance and epidemic outbreak tracking (malaria, HIV/AIDS, TB, Avian Flu, diabetes, ...) as well as diagnostics and treatment support.

4) *Mobile Search:* Mobile search is an evolving branch of information retrieval services that is centred around the

convergence of mobile platforms and mobile devices. It allows users to find mobile content interactively from internet websites as well as enterprise databases.

*5) Mobile Entertainment:* Mobile entertainment comprises a range of activities including but not limited to downloading ring tones, logos, music and movies; playing games, instant messaging, accessing location-based entertainment services, and Internet browsing.

*6) Other Vertical Applications:* Vertical applications include those that address functions and business-specific requirements typically associated with a particular industry or a specific company. These include market segments such as field service, mobile professionals, transportation, point-of-sale, telemetry, governance, and so on.

### III. THE ETHIOPIAN CASE – ETC AND BEYOND

What is the status quo of mobile services in Ethiopia? What enablers are available to be capitalized on?

#### A. Networks, Services and Applications - Now

The sole telecom network and service provider, ETC (Ethiopian Telecommunications Corporation), is currently engaged in large-scale nationwide network expansion projects. The next-generation all-IP network project brings together GSM, CDMA, WCDMA and CDMA2000 mobile networks as well as IP networks and landline PSTN.

ETC plans to reach a staggering 15 million mobile subscribers by 2010. The Chinese ZTE (which is undertaking most of the current projects) and Huawei, Nokia-Siemens, Ericsson, Alcatel-Lucent, Cisco and other telecom network solution providers have been involved with ETC in the past in its telecom network and service provision endeavours for a subscriber base that has been deemed as one of the lowest in the world. [7]

The services are not far beyond basic voice and limited data services. With the expansion of the currently limited 3G network as well as integrated IP networks as a result of the NGN project being finalized, network operators and third party service providers should be able to provide data services (simple text/email messaging as well as internet and value-added services) in addition to basic mobile telephony.

With the advent of mobile device technology, we progressively see handsets with ever-increasing capabilities, like for example, running Java applications or SIM-based software while utilizing content and service from the Telco and others via internet connectivity through packet data (e.g. GPRS or 3G) or alternate bearers like SMS and MMS.

The application developer community and/or ISVs are not found in abundance, for reasons that can directly be associated with the customer base's awareness and the not so conducive network platform and environment for innovation.

As the case is with most technology products and solutions, a compelling mobile service is likely to generate a matching demand of the same scale from customers. This might seem to be a sweeping statement but it generally is true as has been witnessed from experience. One may ask: 'Why has Ethiopian mobile service subscribers not benefited in commensurate ratios with that of, say, Kenya or India?' Is it telecom policy? Is it lack of understanding of how the 'Real Digital Divide' can be bridged? Are we entangled in a vicious circle due to the poor-performing economy? Is the Ethiopian mass unenthusiastic about change via technology adoption at a higher scale compared to other developing countries? Has technology not been transferred enough to the professionals and tech firms in the nation? Can we blame illiteracy?

It's worth mentioning the up-and-coming mobile banking services in Ethiopia where at least half a dozen banks in the country are adopting (or have adopted) e-banking systems. There are projects aimed at providing integrated payment platform with inter-bank cooperation. Mobile banking (SMS-based as well as mobile web) is also on the table. Payment and banking being at the core of most horizontal and vertical mobile services and applications, this is a promising start and we shall expect a lot from the banks and related bodies.

The rather sad report stating that coffee farmers in Ethiopia earn 0.03USD for every 3 USD cup of coffee sold (that's 100 times less!) may at last have been finally resolved. The Ethiopia Commodity Exchange (ECX) is providing IVR (Interactive Voice Response) and SMS-based mobile market information delivery to farmers and traders, avoiding the information gap between the ends of the chain.

Another case was that of UNICEF's effort to coordinate the distribution and maintenance of appropriate food stock in its 1800 feeding centres in the country. The field monitors reported on supplies and number of children fed through an SMS reporting system using a UNICEF-built mobile data collection and monitoring software, RapidSMS.

With all the above being the case, we should seriously consider the effect of illiteracy in the adoption and proper usage of simple to complex mobile services. Nokia's report entitled 'Understanding Non-Literacy as a Barrier to Mobile Phone Communication' [20] discusses the impact illiteracy have had or might have. The launch of mobile handsets with local language support in a number of models by the same company is appreciated. R&D towards easier text entry and better user experience on the mobile phone is essential. Effort to provide mobile operating systems as well as software applications in local languages is expected from device manufacturers as well as software vendors.

#### B. Enablers in the Mobile Arena

We have discussed mobile devices and operating systems in section II. Let's point out wireless application protocols, mobile client software platforms and wireless multimedia technologies as enablers for service and application delivery.

*1) Wireless application protocols:* Include WAP on bearer networks like SMS (Short Message Service), EMS (Enhanced Message Service), MMS (Multimedia Message Service), GPRS (General Packet Radio Service) and mobile IP (IPv4 and IPv6).

*2) Mobile client software technologies:* Mobile web (XML/XSLT/CSS/XHTML), Java 2 Microedition (J2ME),

Flash Lite, .Net Compact Framework and the SIM toolkit are among the commonly used and widely available client software technologies.

*3) Mobile multimedia*: Streaming media architectures (video, flash, and presentation files for example) for mobile networks that provide reasonable QoS have been proposed. [15] Mobile multimedia streaming network protocols defined by standard bodies like 3GPP (Third Generation Partnership Project) and IETF (Internet Engineering Task Force) represent audio/video with standard codecs like MPEG-1, 2, 3 and 4.

*4) Emerged/emerging trends:* Localization (utilizing GPS, Wi-Fi hotspots, GSM Cell Ids, etc.), context-aware computing, semantic web, ad hoc and sensor networking and so on are trends in mobile communications and ubiquitous computing that could enable a range of services to users.

*C. Stakeholders – Who's Involved in What*

There are the telecom network operators (Telcos) which provide the mobile network. Then there are content providers, like media companies. Service providers may be businesses and not-for-profit organizations or the Telcos themselves.

The end users might be individuals or businesses, using mobile handsets from different manufacturers.

Developing, deploying and maintaining the various network services and applications are the technology companies and ISVs.

Providing the manpower and applied/basic research are academic institutions and the academics and researchers in the institutions.

Controlling and enforcing laws and licensing procedures are telecommunications agencies like the ETA (Ethiopian Telecommunications Agency) in Ethiopia or the FCC (Federal Communications Commision) in the USA.

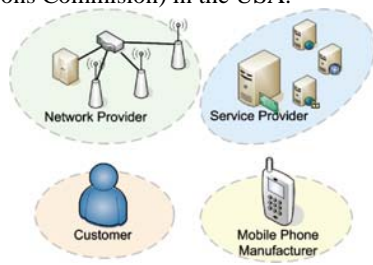

Figure 2 – Players in mobile network services

*D. Open Networks – Parlay/JAIN and Beyond*

Due to the service-based nature of the telecom industry, a recent trend is the so-called Telecom SOA. A service-oriented architecture (SOA) describes an environment where software applications expose functionality via a service provider messaging interface. Other software agents can act as service consumers in using the functionality exposed by service providers. As outlined by the World Wide Web Consortium (W3C), a SOA contains three major components: Discovery, Description and Messaging. [18]

The strength of the telecom world in providing network services to the customer and that of the IT world in offering software applications could be combined to develop new services on top of the underlying network.

The Parlay Group [14], a technical industry consortium founded in 1998 by leaders in the telecom and software industries, specifies APIs for the telephone network which enables the creation of services by organizations both inside and outside the traditional carrier environment. JAIN (Java APIs for Integrated Networks) is the Java version of the Parlay Open APIs.

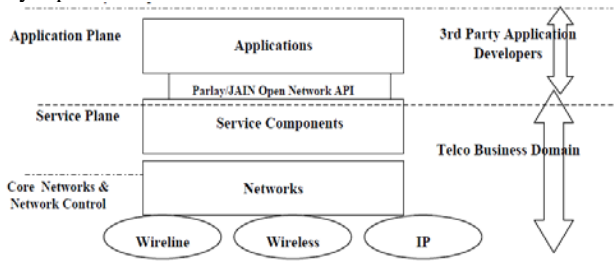

Figure 3 - Open Networks – The New Telecom/IT Paradigm

A typical implementation of Parlay adds a new network element, the Parlay Gateway, on to the underlying (GSM, CDMA or SS7) network. By using JAIN or Parlay X (the web services implementation of the Parlay APIs), the technology allows for Third Party Call, Call Notification, Short messaging, Multimedia Messaging, Payment, Account Management, Terminal Status, Terminal Location, Call Handling, Audio Call, Multimedia Conference, Address List Management, Presence, Message Broadcast, Geocoding, Application Driven QoS, Device Capabilities and Configuration, Multimedia Streaming control and Multimedia Multicast Session Management. [19]

The current ETC network, with NGN softswitch devices provided by ZTE, is known to support the Parlay Framework [16] and can be put to use through open interfaces.

*E. Putting All Together – Mobile Services*

An understanding of the involved industries is required – telecom operators and their business models, content providers/distributers and developers/ISVs in between.

There is an obvious need to understand the customer (both consumers as well as businesses). Things to consider include size, market segmentation, adoption, etc.

How can a winning service offering be created? What makes the mobile service design process different? How to manage the customer experience? What is the role of the mobile devices through which customers access the services? What makes the mobile service environment different? How about mobile service deployment? What do commercial mobile service launches have got to say? How should customer feedback be used for quality improvement?

All of the above are questions that have got to be thoroughly analysed and answered [8] for a successful realization of a mobile service concept.

## IV. THE ROAD AHEAD

What shall be expected from telecom agencies, network operators, businesses and not-for-profit organizations? What about entrepreneurs and tech firms? What's the role played by researchers and academic institutions?

### A. Telecom Agencies

Whilst possible, telecom agencies need to provide a favourable environment for competition in telecom service offerings. This shall be the case even in a monopoly telecom network provision set-up, as service provision and network infrastructure operation should not necessarily mean the same.

Ethiopian Telecommunications Agency (ETA), while overlooking the communications endeavours of the nation, currently licenses what it refers to as 'value added services' - call centres and internet service reselling. This is like chalk and cheese with the real definition of VAS, especially from a mobile telecom perspective.

This being the case, actual execution of the mission of the agency in the direction of its vision needs an acute review of the status quo and experience sharing from corresponding authorities in other countries.

### B. Telecom Network Operators

There is a need to open up networks via APIs such as Parlay/JAIN. This allows for third parties to utilize the network infrastructure for new services.

Formal technical documentation of the network architecture, hardware/software components, changes made to the system, and so on should be available for authorized entities.

Marketing and business development units of Telcos need to devise strategic partnerships and ventures with service providers, content providers like media companies and ISVs to offer new services that would benefit the subscribers while increasing the ARPU (Average Revenue per User).

ETC, specifically, needs to learn from successful Telcos like Safaricom and MTN in the African region plus global giants like Vodafone – both from a technology adoption/implementation perspective as well as innovation in service offerings, corporate VMO/culture and customer satisfaction among others.

### C. Businesses, Government and NGOs

As have been discussed in the experience of other countries, businesses and not-for-profit organizations find numerous use cases beyond everyday mobile telephony for their employees and clientele.

Business-to-business (B2B), business-to-customer (B2C), customer-to-business (C2B) communications on top of mobile phones with custom solutions for a particular business' needs can be enabled with the advent of telecom networks, service offerings and mobile applications.

A grassroots NGO educating the youth via mobile messaging, a research institution collecting data through mobile phones, a business promoting its service and whereabouts using location-based mobile advertising are few scenarios that can be thought of.

The bottom line is that businesses and organizations need to devise ways to tap the potential mobile services offer. A list of 'appropriate technologies' for a business/organization shall now include the mobile network.

### D. Entrepreneurs and Innovators

The mobile economy endows entrepreneurs and innovators with a canvas for creativity and promising markets. With a thorough understanding of the principles behind the engineering of mobile services & applications, innovation comes in the rescue for several development problems.

A compelling effort is an initiative led by Nathan Eagle at Massachusetts Institute of Technology (MIT). The EPROM (Entrepreneurial Programming and Research on Mobiles) initiative works with academics and start-up companies in sub-Saharan Africa, mainly Kenya and Uganda but with presence in other counties like Ethiopia as well. [13]

### E. Technology Companies and Professionals

MobileBeat, a yearly event that showcases start-ups and ventures in the up-and-coming mobile economy, features remarkable companies and people working on the cutting-edge of mobile communications and computing. [17]

Mobile software applications and services, location based systems, user experience, multimedia technologies, security, mobile middleware, near field communications, application specific chips, and so on are what tech firms in the making are being engaged on. Professionals can see prospects in applied R&D in the aforementioned areas.

### F. Researchers and Academic Institutions

The research and development of mobile network services and applications is truly multidisciplinary. It involves, among others, people with a background in socio-economics, telecom management and engineering, business, marketing, computer science and engineering. In addition, experts from the domain in question (say, health or agriculture) to which the mobile solution is being used for will be needed.

Fisseha Mekuria [12] describes a program for mobile computing and applications software development (MCASD), in a developing region perspective, having three 'pillars' - a master's program; a mobile lab for development, prototyping and usability testing; and research in mobile technology, content and services.

In cases where a separate program may not appeal for academic institutions, seminars and/or inclusions in course lectures related to communication systems, operating systems and software development/engineering may suffice, at least in technical institutions.

Researchers affiliated with research groups may conduct community-based applied research and development in mobile networks, services and applications while strengthening industry-academic relationships.

The role played by the academics/researchers as well as the academic institutions hosting students in technology and the humanities can't be overemphasized. Lessons can be learnt from Europe, for example, where telecom operators like BT and network solution providers like Nokia partner with academic institutions on different R&D projects.

ETC's Graduate School for Telecommunications and Information Technology (GSTIT) has been one good start in Ethiopia but is short of what can be achieved. Something more than a school run by the Telco itself is obviously needed.

A centre of excellence by the government, the mobile network operators, service providers, academic institutions and other stakeholders seems to be a viable answer to create a platform for innovation towards apposite solutions to the nation's endeavours in agriculture, health, education, business, and so on, via the now ubiquitous mobile phone.

## V. CONCLUSIONS

This paper described the role mobile services can play in the networked economy touching upon notions like ICT4D and the 'digital divide'. Mobile phones and network services beyond voice communication are tipped to drive growth in poor nations like Ethiopia. In addition, the still broad rich-poor 'digital divide' can be bridged via the more readily available mobile networks, services as well as now ubiquitous mobile handsets.

The paper pointed out what is expected of stakeholders in mobile telecom technology, policy and business arenas. The need to learn from the experience of successful mobile service launches in diverse sectors (like business, finance, healthcare, governance, education and infotainment) in different countries has been stressed bearing in mind the economic and social impact they have had.

Technical overview of mobile service engineering has been described, underlining the need for open networks via Parlay/OSA and outlining some enabling (emerged/emerging) technologies that can be put to use such as wireless application protocols, client software technologies, mobile multimedia and trends in localization, context-aware computing, semantic web, et cetera.

ICTs could play a vital role in a nation's development endeavors and mobile phones, coupled with appropriate network services, could contribute the lion's share in ICT4D. Key players in Ethiopia, the theme of this paper, namely the government, ETA, ETC, tech firms and academic institutions, etc., need to respond promptly with action, joint initiatives and strategies to tap the 'huge potential at hand', literally.